\documentstyle[preprint,aps]{revtex}

\begin{document}
\draft
\preprint{}
\title{Comment on the ``Chaotic'' Singularity in Some Magnetic Bianchi
VI$_0$ Cosmologies \thanks{Permanent address:
Physics Department, Oakland University, Rochester, MI  48309, USA.
\protect \newline e-mail: berger@oakland.edu}}
\author{Beverly K.~Berger}
\address{Albert Einstein Institute,
Max-Planck-Institut f\"{u}r Gravitationphysik,\\
Schlaatzweg 1, D-14473 Potsdam, Germany}
\date{\today}
\maketitle
\pacs{98.80.Dr, 04.20.Jb}

\narrowtext

\begin{abstract}
Description of the magnetic Bianchi VI$_0$ cosmologies of LeBlanc,
Kerr, and Wainwright in the formalisms both of Belinskii, Khalatnikov, and
Lifshitz, and of Misner allows qualitative understanding of the
Mixmaster-like singularity in those models.
\end{abstract}

\newpage

LeBlanc, Kerr, and Wainwright (LKW) \cite{LKW} have recently used the
dynamical systems approach popularized by Wainwright and Hsu \cite{HW}
to study the class of Bianchi VI$_0$ cosmologies containing a magnetic
field and a perfect fluid. As part of their complete analysis of this
system, they reported that the generic singularity (for most fluid
equations of state) was of the
Mixmaster type \cite{BKL,Misner}. The purpose of this comment is to
elucidate the nature of this feature of the models by reexpressing the
results in the Belinskii, Khalatnikov, and Lifshitz (BKL) and Misner's
minisuperspace (MSS)
terminologies. While nothing new is thereby revealed, those more conversant
with these formalisms will find LKW's somewhat surprising result less
mysterious.
In most of the following, we shall restrict attention to the electrovac case,
only remarking at the end on the effect of inclusion of a perfect fluid.

First consider the ``standard Mixmaster model'' (SMM) of (diagonal)
vacuum, Bianchi IX. The (BKL) metric components $a^2$, $b^2$, and $c^2$ are
related
to the MSS variables $\Omega$ and $\beta_{\pm}$ describing volume and
anisotropy by
\begin{eqnarray}
\label{coords}
\alpha &=& \ln a \ = \ \Omega -2 \beta_+ , \nonumber \\
\zeta &=& \ln b\  =\  \Omega + \beta_+ + \sqrt{3} \beta_- , \nonumber \\
\gamma &=& \ln c \ = \ \Omega + \beta_+ - \sqrt{3} \beta_- .
\end{eqnarray}
Einstein's equations may be obtained by variation of the superhamiltonian
${\tilde{\cal N}}{\cal H}$\footnote{The rescaled lapse is $\tilde{{\cal N}}
= {\cal N} e^{- 3
\Omega}$ where $dt = {\cal N} d \tau$ with $\tau$ our choice of time
coordinate and $t$ comoving proper time. In LKW, $\cdot \equiv d/{dt}$
while $\tilde{\cal N} = 1$ yields the BKL time coordinate.} with
\begin{equation}
\label{h0mss}
2 {\cal H} = - p_{\Omega}^2 + p_+^2 + p_-^2 +
U(\Omega,\beta_{\pm})
\end{equation}
where $p_{\Omega}$ and $p_{\pm}$ are respectively canonically conjugate
to $\Omega$ and $\beta_{\pm}$ and
\begin{eqnarray}
\label{Umss}
U &=& e^{4 \Omega} V(\beta_+,\beta_-) \nonumber \\
 &=& e^{4 \alpha} + e^{4 \zeta} + e^{4 \gamma} -2 e^{2(\zeta + \gamma)}
-2 e^{2(\alpha + \gamma)} -2 e^{2 (\alpha + \zeta)}.
\end{eqnarray}
Arbitrary constants in front of $U$ may be absorbed by a suitable choice
of spatial coordinates. In (\ref{h0mss}), the MSS potential, $U$, is related
to the spatial scalar curvature through $U = e^{6 \Omega} \ \ ^3\/R$ where
$e^{6 \Omega}$ is the determinant of the spatial metric with scalar curvature
$\ ^3\/R$.
Direct comparison of (\ref{h0mss}) and (\ref{Umss}) with the LKW form
of the Hamiltonian constraint suggests the identification
\begin{equation}
\label{definesigmaN}
\Sigma_{\pm} = {{p_{\pm}} \over {- p_{\Omega}}} \quad ; \quad
N_1 = c_1 {{e^{2 \alpha}} \over {-p_{\Omega}}} \quad ,
\end{equation}
etc. LKW's equations of motion are then obtained with $\tilde{\cal N} =
-p_{\Omega}^{-1}$ so that the time coordinate is $\Omega$. The
$c_i$ contain
factors from the structure constants. Thus change of Bianchi
type can lead to vanishing of some of the $N_i$ and sign changes of
others \cite{HW}.

The approach to the singularity ($\Omega \to -\infty$)
in the SMM has long been known to be characterized by an infinite sequence
of Kasner models (vacuum Bianchi I) described by the single BKL parameter
$u$ \cite{BKL,Misner}.
The Kasner solution is just that obtained for $U = 0$ in (\ref{h0mss}).
The superhamiltonian becomes that for
a free particle so that the momenta $p_{\Omega}$ and $p_{\pm}$ are constant.
The Kasner epoch changes when a scattering (bounce) off the MSS potential $U$
occurs.\footnote{The overall $e^{4 \Omega}$ dependence of $U$ means that, as
$\Omega \to - \infty$, $U \to 0$. However, the Kasner solution (expressible
as $\beta_{\pm}= \beta_{\pm}^0 + \Sigma_{\pm}(\Omega - \Omega_0)$) allows
terms in $U$ to be of order unity (e.g.~if $\Omega = 2 \beta_+$).} The values
of $u$ in successive Kasner eras are related by
the BKL map \cite{BKL}, equivalent to the Kasner map described
by LKW \cite{LKW}. All discussions of chaos in the SMM derive from the
well-known sensitivity to initial conditions of this map (see for example
\cite{hobill}). Here we need consider only the features of the SMM which are
required to derive the BKL (or Kasner) map since the map will result in any
model in which these same properties are present. In fact, only the first
three terms in $U$ are needed.\footnote{One can argue that
the remaining terms are always dominated by one of the first three except
near the $120^{\circ}$ corners of the MSS potential \cite{Misner}.
In this discussion of
generic Mixmaster behavior, the ``anomalous'' corner behavior will be
ignored. See, however, \cite{BKL,MW}.} Since the Kasner solution is
characterized (in the approach toward the singularity) by two contracting
metric components and one expanding metric component, one of $\alpha$,
$\zeta$,
or $\gamma$ will dominate. For convenience, choose the dominant one to be
$\alpha$ so that $U \approx e^{4 \Omega - 8 \beta_+}$. The canonical
transformation generated by $x = \Omega - 2 \beta_+$ and $y = - 2 \Omega +
\beta_+$
puts the Hamiltonian constraint (\ref{h0mss}) into the form
\begin{equation}
\label{h0II}
3 p_x^2 - 3 p_y^2 + p_-^2 + e^{4 x} = 0.
\end{equation}
Eq.~(\ref{h0II}) describes scattering off a potential with relationships
among the momenta before and after the scattering of constant $p_y$ and
$p_-$ with $p_x \to - p_x$.
Since the BKL parameter $u$ may be expressed in terms
of these momenta \cite{BKL}, the scattering rules lead immediately to
the BKL map. It is important to note that, even though the exact
solution corresponding to (\ref{h0II}) is known, only the momentum
rules are used.

With the identifications (\ref{definesigmaN}), the Hamiltonian constraint
yields the identity
\begin{equation}
\label{useconstraint}
p_{\Omega}^{-2} U = 1 - \Sigma
\end{equation}
(where $\Sigma = \Sigma_+^2 + \Sigma_-^2$ \cite{LKW}) so that
\begin{equation}
\label{lnpweq}
\frac{d}{d \Omega} \ln p_{\Omega} = 2 - q
\end{equation}
where $q = 2 \Sigma$. The LKW equations for the $N_i$ then yield the
relationship of these variables to the MSS or BKL ones.
For example, $N_1 = {n_1} / {p_{\Omega}}$ in \cite{LKW}
\begin{equation}
\label{n1eq}
\frac{d N_1}{d \Omega} = (q - 4 \Sigma_+ ) N_1
\end{equation}
with (\ref{lnpweq}) has the solution (using $d \beta_{\pm}/d \Omega
= \Sigma_{\pm}$)
\begin{equation}
\label{n1form}
n_1 = n_1^0 e^{2 \Omega - 4 \beta_+}.
\end{equation}
Similar constructions can be performed for
$N_2$ and $N_3$.

The most attractive feature of the LKW formalism is its ability to handle all
the Bianchi types with the same equations. Thus extension to the magnetic
(but otherwise empty) Bianchi VI$_0$ equations is straightforward.
To make the
connection to the BKL and MSS formulations, however, we must include some
properties of the structure constants (i.e. in the $c_i$ of
(\ref{definesigmaN})).
Thus, for Bianchi VI$_0$, $c_1 = 0$ and $c_2 c_3 < 0$ while
$H_1 \equiv {h_1} / {p_{\Omega}}$ is the only magnetic field component
consistent with this model \cite{LKW}. The magnetic field contribution to the
Hamiltonian constraint leads to a modification of (\ref{useconstraint}) such
that $p_{\Omega}^{-2} U = 1 - \Sigma - \frac{3}{2}H_1^2$ and $q = 2
\Sigma + \frac{3}{2} H_1^2$.
These identifications allow
the LKW equation for $H_1$ \cite{LKW}
\begin{equation}
\label{H1eq}
\frac{d H_1}{d \Omega} = (q - 1 - 2 \Sigma_+) H_1
\end{equation}
to yield the solution
\begin{equation}
\label{h1solution}
h_1 = h_1^0 e^{\Omega - 2 \beta_+}.
\end{equation}
It is now possible to transcribe the LKW constraint for this model
into the MSS language as
\begin{equation}
\label{h0vi}
2 {\cal H} = - p_{\Omega}^2 + p_+^2 + p_-^2 + \left[ e^{4 \zeta}
 +  e^{4 \gamma}
 + 2 e^{2 (\zeta + \gamma)}\right]
+ \xi e^{2 \alpha}.
\end{equation}
We note that, compared to the SMM constraint (\ref{h0mss}),
the Kasner (momentum) part is unchanged.
The first two terms in the brackets, identical to those in SMM,
are two of those required for
the BKL map derivation. The third term is the
irrelevant cross term (with irrelevant changed sign). The final term
in (\ref{h0vi}), from the magnetic field, is
essentially the square root of the corresponding term in the SMM potential.
However, the arbitrary constants, the absent cross
terms, and the difference in the power of $e^{\alpha}$ have no effect on
the BKL map derivation. The same canonical transformation to $x$ and $y$
can be made yielding a model for the scattering from one Kasner epoch to
another of
\begin{equation}
\label{h0H}
3 p_x^2 - 3 p_y^2 + p_-^2 + \xi e^{2x} = 0
\end{equation}
rather than (\ref{h0II}). Even though (as pointed out by LKW) the exact
solution has
changed, the rules describing the change in momenta at the bounce have
not.\footnote{Of course, (\ref{h0II}) is exactly reproduced to describe
bounces for either $\zeta$ or $\gamma$ approximately zero.}
Thus the BKL map may be derived as a property of the magnetic Bianchi VI$_0$
vacuum model.

Finally, we note that a perfect fluid will enter the Hamiltonian constraint
(\ref{h0mss}) as
a term $U \to U + \rho_0 e^{3 (2 - \Gamma) \Omega}$ for equation of state
$p = (\Gamma -1) \rho$. As $\Omega \to - \infty$, this term goes to zero
(if $\Gamma \ne 2$ in the range discussed by LKW) and can never revive due
to cancellations such as those in other terms of $U$
that produce Mixmaster bounces. Only if $\Gamma = 2$ will the fluid play a
role in the singularity approach (as has been found by LKW).

Thus, the connection between the LKW formalism and the BKL and MSS
descriptions
of Bianchi cosmologies clarifies for those familiar with the latter the
similarity between the singularity dynamics of the SMM and magnetic Bianchi
VI$_0$ models by showing that the differences between the two fail to
invalidate the derivation of the BKL map. Of course, the BKL approximate
description of the SMM contains more than just the parameter $u$
\cite{BKL,CB}. While the exact solution to the bounce model equation is
still not needed for this expanded description, differences between
the two models might prove interesting.

\section*{Acknowledgments}
I would like to thank Alan Rendall for useful comments.
This work was supported in part by National Science Foundation
Grant PHY9507313.

\end{document}